\documentclass[twocolumn,showpacs,preprintnumbers,amsmath,amssymb]{revtex4}

\usepackage{graphicx}
\usepackage{amsthm}
\usepackage{epsfig}   
\usepackage{amssymb}
\usepackage{color}
\usepackage{graphicx}  
\usepackage{dcolumn}   
\usepackage{bm}

\input{epsf}

\def\be{\begin{equation}}
\def\bea{\begin{eqnarray}}
\def\ee{\end{equation}}
\def\eea{\end{eqnarray}}

\def\apx{\approx}

\def\al{\alpha}

\def\gm{\gamma}
\def\Gm{\Gamma}

\def\Dt{\Delta}
\def\eps{\varepsilon}

\def\Om{\Omega}
\def\om{\omega}

\def\dd{\mbox{d}}

\def\mod{\mbox{ mod}}

\begin{document}

\preprint{APS/123-QED}

\title{Tailored mixing inside a translating droplet\\}

\author{R. Chabreyrie$^1$, D. Vainchtein$^{2,3}$, C. Chandre$^4$, P. Singh$^5$, N. Aubry$^1$}
\affiliation{$^1$ Mechanical Engineering Department, Carnegie Mellon
University, PA 15213, USA \\ $^2$ School of Physics, Georgia
Institute of Technology, GA 30332, USA \\ $^3$Space Research
Institute, Moscow, GSP-7, 117997, Russia \\$^4$ Centre de Physique
Th\'eorique, Luminy-case 907, F-13288 Marseille cedex 09, France \\
$^5$ Mechanical Engineering Department, New Jersey Institute of
Technology, Newark, NJ 07102, USA}

\date{\today}

\begin{abstract}
Tailored mixing inside individual droplets could be useful to ensure
that reactions within microscopic discrete fluid volumes, which are
used as microreactors in ``digital microfluidic'' applications, take
place in a controlled fashion. In this article we consider a
translating spherical liquid drop to which we impose a time periodic
rigid-body rotation.  Such a rotation not only induces mixing via
chaotic advection, which operates through the stretching and folding
of material lines, but also offers the possibility of tuning the
mixing by controlling the location and size of the mixing region. Tuned mixing is  achieved
by judiciously adjusting the amplitude and frequency of the
rotation, which are determined by using a resonance condition and
following the evolution of adiabatic invariants. As the size of the mixing region is increased,
complete mixing within the drop is obtained. 
\end{abstract}

                                           
\maketitle

\section{\label{sec:intro}Introduction}
Droplets have been proposed as an alternative to standard
fluid-stream microfluidics for lab-on-a-chip applications. This
microfluidics approach, also referred to as ``digital'' because it
uses ``discrete'' fluid volumes (droplets) rather than continuous
streams, holds great promise due to the possibility of using single
droplets as microreactors \cite{Ismagilov:2003}. Efficient mixing,
however, is needed for reactions to occur, but remains difficult to
achieve because the Reynolds number (Re) is usually very small and so
the flow is laminar. This issue has recently attracted much attention
in the literature.  For flows in microchannels, while there are many
strategies based on altering the channel geometry, the use of
forcing alone (see, e.g., \cite{Oddy:2001,
Bau:2001, Ouldelmoctar:2003, GlasgowAubry:2003, Glasgow:2004}) has also proved to be efficient,
especially in the case of low Re \cite{Goullet:2006}.  The
combination of both geometry alteration and forcing has been
explored as well \cite{Niu:2003, Bottausci:2004, Stremler:2004,
Goullet:2006}.  For droplet-based microfluidics, the forcing alone
is the preferred strategy as the deformation of the droplet is
difficult to control. In almost all cases, the enhancement of mixing
in miniature geometries is based on chaotic
advection, the stirring phenomenon that stretches and folds fluid
elements thus increasing the interfacial area
between the two fluids to be mixed.
Chaotic advection inside a liquid drop subjected to a forcing (at low Re) has been studied extensively \cite{Baj, KroujilineandStone:1999, Lee:2000, WardandHomsy:2001, 
Grigoriev:2005, Homsy:2007,
VWG:2007}  and was obtained experimentally by means of oscillatory flows \cite{WardandHomsy:2003,GSS:2006}.
In this letter, we focus on unsteady -- yet periodic --
forcing.\\
From a dynamical systems viewpoint, the introduction of a time-dependent perturbation or forcing breaks the
invariants (related to the symmetries of the unperturbed system), thus introducing resonances between the natural
frequencies of the unperturbed problem and the frequency(-ies) of the
forcing. Although such resonances create chaotic regions where mixing occurs, in general, chaotic and regular regions co-exist  and unexpected regular sizable pockets persist.\\
In many situations where it is indeed possible to create chaos,
controlling the mixing region(s) remains a challenge. Such a
control, however, should be possible since a chaotic system is
sensitive to changes in parameter values (as it is to changes in
initial conditions). These changes should generically modify the
resonances, and thus the location and size of the chaotic regions.\\
Our general approach along these lines is to consider a bounded
three-dimensional (3D) flow, which is the superposition of an
integrable flow ${\bf v}_0$ with at least one invariant  and
a small time-dependent perturbation $\eps {\bf v}_1({\bf x},t)$,
$0\le \eps \ll 1$. If ${\bf v}_0$ has only one invariant, the phase
space contains two-dimensional  tori. In this
case, the perturbed flow, $\dot{\bf x}={\bf v}_0({\bf x})+\eps {\bf
v}_1({\bf x},t)$, has poor mixing properties if the amplitude of the
perturbation $\eps$ is small, since two-dimensional (2D) tori act as
barriers to chaotic diffusion (e.g., \cite{FKP:1988}). If, on
the other hand, ${\bf v}_0$ has two invariants, trajectories  of
this integrable flow are all periodic. Most of these periodic orbits
are expected to be broken by a generic perturbation ${\bf v}_1$ with
an arbitrarily small amplitude $\eps$. Efficient mixing properties
might then be obtained with such perturbed flows. In this work, we
consider an axisymmetric integrable flow possessing two invariants,
thus possibly offering efficient mixing properties after being
perturbed.\\
While many previous works \cite{Baj, KroujilineandStone:1999,
VVN:1996a, VNM:2006} have shown the existence of chaotic behavior in
3D bounded steady flows, we turn our attention to unsteady flows;
the added unsteadiness targets the control of the chaotic behavior
through resonance phenomena \cite{Lima:1990, CFP2:1996, VWG:2007}.  Specifically, we seek to create a mixing zone of tunable size which remains localized within a well-defined region of the drop.  This should also provide a rationale for the route to complete mixing as the perturbation increases.
\section{\label{sec:conclu} Model}
\subsection{\label{sec:conclu} Flow equation and assumptions}
We consider a spherical Newtonian drop immersed in an
incompressible Newtonian flow in the case where the linear external
field is characterized by translational velocity and vorticity
vectors, similarly to \cite{KroujilineandStone:1999}.  As in the
latter reference, we assume that the local Re is much
smaller than one and that the interfacial tension is sufficiently
large for the drop to remain spherical.\\
The internal velocity field is obtained by solving the Stokes flow
problem for both the internal and external flows satisfying the
continuity of velocity and tangential stress conditions across the
drop surface. In addition, we introduce unsteadiness in the problem
by making the vorticity time dependent. In a Cartesian coordinate
system translating with the center-of-mass velocity of the drop, and
with the $z$ axis in the direction of the translation, the
 paths of passive marker particles are given by the solution of the non-autonomous
dynamical system:

\begin{eqnarray}
\nonumber
u=\dot{x}&=& zx -  a(t) \om_z y,\\
\label{veloT22}
v=\dot{y}&=& zy +  a(t)\left(\om_z x - \om_x z\right),\\
\nonumber
w=\dot{z}&=& (1-2x^2-2y^2-z^2) +  a(t) \om_x y,
\end{eqnarray}

where all lengths and velocities have been non-dimensionalized by
the drop radius and the magnitude of the translational velocity.
Here, the vorticity is defined by $\bm{\om} =
(\om_x,\om_y,\om_z)=(1/\sqrt{2},0,1/\sqrt{2})$, the unitary vector
corresponding to the axis of rotation, and $\label{pert}
a(t)=\eps/2\left(1+\cos \om t \right)$, characterized by the
frequency  $\om$ and the amplitude $\eps$. In this letter, we
consider only small amplitudes, i.e.\ for $0 \le \eps \ll 1$.  Note
that the former equations are identical to those in
\cite{KroujilineandStone:1999} except that  the constant vorticity vector
has been replaced by $a(t)\bm{\om}$. This can be done by either
assuming unsteady vorticity in the external flow field, or by
applying a time dependent body force. In practice, this could be
realized, e.g., by creating a time dependent swirl motion in the
external flow or by applying an electric field that exerts a torque
on the drop (e.g.,\cite{AubrySingh:2006} it or work on electrorotation). This flow is the
superposition of a Hill's vortex and an unsteady rigid body
rotation, and the surface of the drop, $r^2 = x^2+y^2+z^2 =1$, is
invariant under  flow (\ref{veloT22}).
\subsection{\label{subsec:concluI} Integrable case}
\begin{figure}[t]
\center\epsfig{file=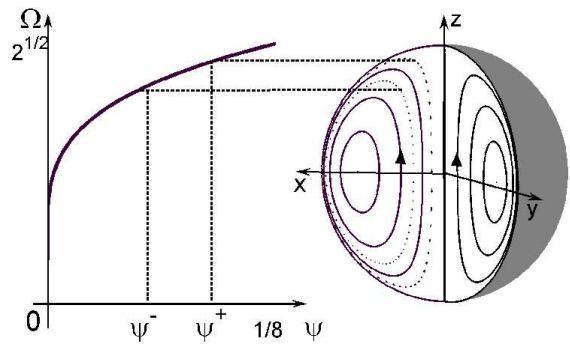, width=3in}
\caption{\label{structure} Streamlines inside the drop (without
rotation) and  their frequencies $\Om\left(\psi\right)$ as given by Eq.~(\ref{Per}).
}
\end{figure}

We now discuss some features of the unperturbed axisymmetric (2D)
flow ($\eps =0$). The flow possesses two independent integrals of
motion, e.g., the streamfunction $\psi$ and the azimuthal angle
$\phi$:
\be
\psi = 1/2 \rho^2 \left(1-r^2\right), \quad \phi =\arctan y/x,
\ee
where $\rho^2=x^2+y^2$ and $\psi \in \left[0,1/8\right]$. The
streamlines of the unperturbed system are  lines of constant $\psi$
and $\phi$, denoted by $\Gm_{\psi,\phi}$, and defined as
$(1-2\rho^2)^2+(2\rho z)^2 = 1-8\psi$ (see Fig.~\ref{structure}).
Almost all streamlines are closed curves surrounding a circle of degenerate
elliptic fixed points $(\rho=1/\sqrt{2},z=0)$. In addition, there
are two hyperbolic fixed points located at the poles of the sphere
which are connected by heteroclinic orbits. The frequency of the motion on
$\Gm_{\psi,\phi}$ is given by
\be
\frac{2\pi}{\Om(\psi)} = \int^{\pi/2}_{-\pi/2}\frac{\sqrt{2} \;
\dd\al}{\sqrt{1+\gm(\psi)\sin\al}} =\frac{2\sqrt{2}}{\sqrt{1+\gm}}
K\left( \sqrt{\frac{2\gm}{1+\gm}}\right),
\label{Per}
\ee
where $\gm(\psi)=\sqrt{1-8\psi}$ and $K$ is the complete elliptic
function of the first kind. The frequency $\Om$ is bounded by two
limits, $\Om(0)=0$ and
$\Om(1/8)=\sqrt{2}$ (see Fig.~\ref{structure}).

On every streamline $\Gm_{\psi,\phi}$, we introduce a uniform phase
$\chi~\mod (2\pi)$ such that $\chi=0$ on the $x-y$ plane (with $\rho
\le 1/\sqrt{2}$) and $\dot{\chi}=\Om\left(\psi\right)$. The
unperturbed system, which can be rewritten in terms of
$(\psi,\phi,\chi)$ as
\[
\dot{\psi} =0, \quad \dot{\phi} =0, \quad \dot{\chi} = \Om(\psi),
\]
belongs to the class of action-action-angle flows.
\subsection{\label{subsec:modelII} Perturbed case}
In the perturbed case $0<\eps\ll 1$, the time evolution of the two
invariants of the unperturbed system is given by
\begin{eqnarray}
\nonumber
\dot{\psi}&=&-2 a(t)\om_x\psi\sin\phi G\left(\psi,\chi\right),\\
\label{psiphi}
\dot{\phi}&=&a(t)\om_z - a(t)\om_x\cos\phi G\left(\psi,\chi\right),
\end{eqnarray}
where $G(\psi,\chi)=z/\rho$ is $2\pi$ periodic in $\chi$ and has
zero average in $\chi$. The time evolution equation for $\chi$ is
\be
\nonumber
\dot{\chi}=\Om(\psi)+a(t) H(\psi,\phi, \chi),
\ee
where $H$ is $2\pi$ periodic in $\chi$. The dynamics possesses two
 time scales, a fast one (of order one) associated with
$\chi$, and a slow one (of order $1/\eps$) associated with $\psi$
and $\phi$.\\
If $\Om$ and $\om$ are incommensurate, then the averaging over $\Om$
and over $\om$ can be performed independently. In this case, the
time-periodic terms in Eq.~(\ref{psiphi}) average out, and the {\em
averaged system} reduces to $ \dot{\psi} = 0 , \quad \dot{\phi} = -
\eps/2$. Thus in the averaged system the value of $\psi$ is
conserved as it was in the unperturbed system; in other words,
$\psi$ is an invariant of the averaged system. Each trajectory of
the averaged system evolves on two-dimensional nested tori
${\mathcal T}_\psi$. In the perturbed system, $\psi$ is an adiabatic
invariant and the motion follows adiabatically the tori ${\mathcal
T}_\psi$.

\section{\label{sec:resul}Methods and results}
\subsection{\label{sec:resul}Mixing generation via resonance phenomena}
We now turn to the generation of a 3D chaotic
mixing region inside the drop, for which we seek to
control both the location and the size. The strategy used for
this purpose is to bring a chosen family of unperturbed tori
${\mathcal T}_{\psi}$ into resonance with the perturbation $a(t)$ by
adjusting the frequency $\om$ to satisfy the resonance condition
\be
\label{resonance}
n\Om(\psi) - \om =0,
\ee
for some $n \in \mathbb{N}$ (see Fig.~\ref{structure}). For any
fixed $\om$ we denote by $\left\{{\mathcal T}^{(n)}(\om)~|n
\in \mathbb{N}\right\}$ the set of resonant tori ${\mathcal T}_{\psi}$ satisfying
(\ref{resonance}). Hereafter, we denote  the chaotic mixing
region generated around ${\mathcal T}^{(1)}(\om)$ by CMR.
\subsection{\label{sec:resul}Control of the mixing}

\begin{figure}[t]
\center\epsfig{file=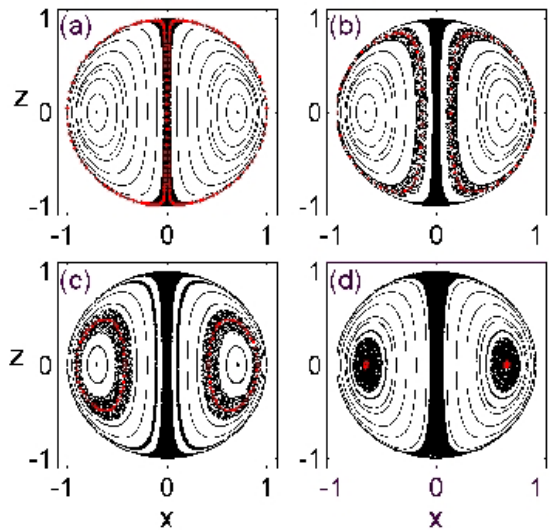, width=3in} \caption{\label{f2}
Liouvillian sections for the amplitude $\eps=0.03$ and the frequencies  $\om=0.55, 0.93,1.28,1.41$ (a-d). The (red) dashed line inside
the CMR is the torus ${\mathcal T}^{(1)}$.}
\end{figure}

\begin{figure}[t]
\center\epsfig{file=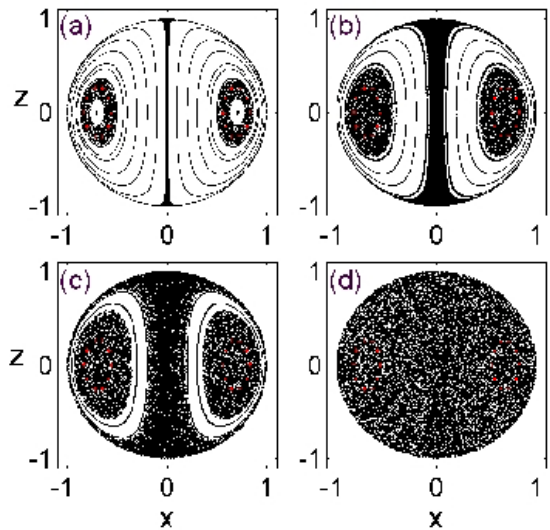, width=3in} \caption{\label{sizeII}
Liouvillian sections for the frequency $\om=1.376$ and the amplitudes  $\eps=0.01,0.05,0.10,0.20$ (a-d).}
\end{figure}

Figures~\ref{f2} and ~\ref{sizeII} present {\it Liouvillian
sections} of the perturbed system, which consist of 2D projections
of time-periodic 3D flows by a combination of a stroboscopic map
and a Poincar\'e section (here, the $y=0$ plane). Figure~\ref{f2}
shows that a perturbation $a(t)$ creates a 3D CMR around ${\mathcal
T}^{(1)}(\om)$ and its location is controlled by varying $\om$
according to Eq.~(\ref{resonance}). In what follows, we analyze the
location and the size of the CMR as $\om$ and $\eps$ vary.

\begin{figure}[htbp]
\center\epsfig{file=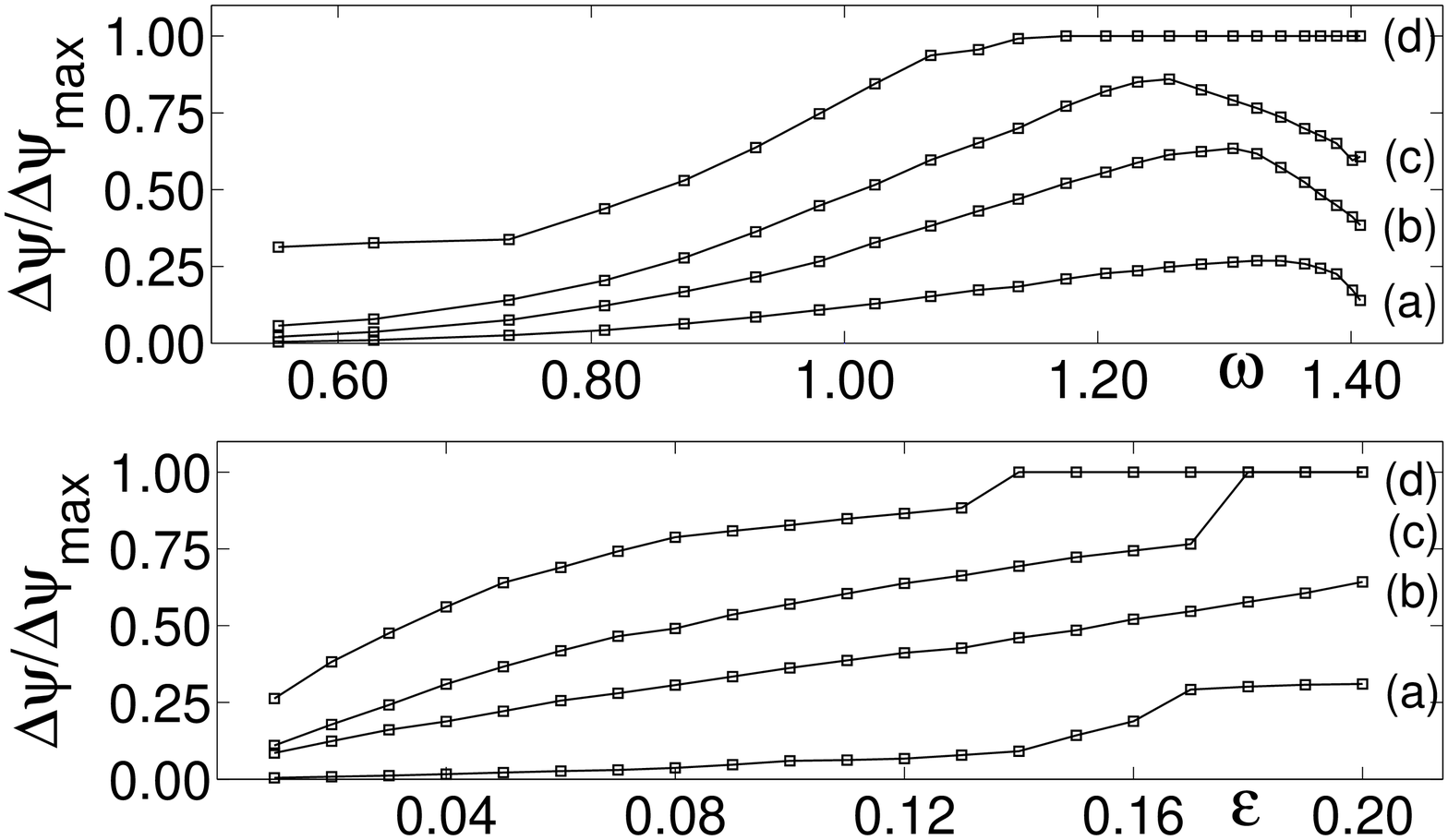, width=3in} \caption{\label{size}
Size of the chaotic mixing region; Upper panel: Normalized $\Dt\psi$ {\it vs.}
$\om$ for the amplitudes $\eps=0.01,0.05,0.10,0.20$ (a-d); Lower panel: Normalized $\Dt\psi$ {\it vs.} $\eps$ for  $\om=0.55,0.93,1.28,1.41$ (a-d).}
\end{figure}

For small values of $\om$, all resonances are located near the
pole-to-pole heteroclinic connections (at $\psi=0$, near the $z$
axis and near the boundaries of the drop, see Fig.~\ref{f2}a). As
$\om$ is increased, the CMR penetrates deeper into the drop
(Fig.~\ref{f2}b).  In the interval $0<\om< \sqrt{2}$, the CMR is the
largest chaotic region (compared to chaotic regions corresponding to
higher order resonances), with all the other chaotic regions
localized close to the $z$ axis and near the drop boundaries (around
the heteroclinic orbits); this is due to the shape of
$\Om\left(\psi\right)$.  As $\om$ is increased further, the CMR
moves toward the location of the elliptic fixed points of the
unperturbed system, closely following the location of the
resonant torus ${\mathcal T}^{(1)}(\om)$ (Fig.~\ref{f2}c). As the
value of $\om$ approaches $\sqrt{2}$, the CMR  shrinks to the circle
of elliptic fixed points (Fig.~\ref{size}).

Whereas the frequency $\om$ of the rigid body rotation is mostly
responsible for the location of the CMR, it is its amplitude $\eps$
which mostly determines its size. Figure~\ref{sizeII} shows that the
size of the chaotic mixing regions created by the $n=1$ resonance
and by higher order resonances (mostly the $n=2$ resonance)
increases as the amplitude of the perturbation increases. Around
$\eps\apx 0.20$, the chaotic regions around the heteroclinic orbits
and the CMR join together to cover the entire drop volume.

\begin{figure}[htbp]
\center\epsfig{file=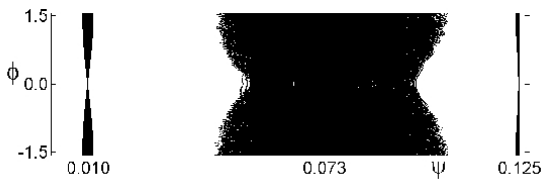, width=3in} \caption{\label{order}
Projection of three characteristic trajectories on the slow phase
plane, with $\phi_0=0$ and $\psi_0=0.010,0.073,0.125$.}
\end{figure}

Recall that in the averaged system the adiabatic invariant $\psi$ is
constant. In the exact system, however, along a given trajectory
starting at $\psi=\psi_0$ it varies between $\psi^-\left(\psi_0;
\om,\eps \right)$ and $\psi^+\left(\psi_0; \om,\eps \right)$. The
width $\Dt \psi= \psi^+\left(\psi_0; \om,\eps\right)-
\psi^-\left(\psi_0; \om,\eps \right)$ is small away from the
resonance, and increases significantly closer to the resonance. The
projection of three characteristic trajectories onto the
$(\psi,\phi)$-plane (called the {\it slow plane} in 
dynamical systems) is presented in Fig.~\ref{order}. The narrow regions on the sides are off-resonance
trajectories that stay quite close to the corresponding tori
${\mathcal T}_{\psi}$. In between, the middle trajectory deviates
much further from its ${\mathcal T}_{\psi}={\mathcal T}^{(1)}(\om)$
and fills the entire CMR. The quantity $\Dt \psi$ is probably the
most convenient quantity to estimate the size of the
CMR (around ${\mathcal T}^{(1)}(\om)$ for $\om < \sqrt{2}$). The
volume between the tori ${\mathcal T}_{\psi^-}(\om)$ and ${\mathcal
T}_{\psi^+}(\om)$ gives the CMR size in 3D.\\
The dependence of the size of the CMR (in terms of $\Dt \psi$) on
$\eps$ and $\om$ is illustrated in Fig.~\ref{size}. The curves
(a)-(d) in the upper and lower panels correspond to
the Figs.~\ref{f2}a-d and ~\ref{sizeII}a-d,
respectively. For a given $\om$ value (i.e. for a given 
${\mathcal T}^{(1)}(\om)$), the size can be controlled by adjusting
the value of $\eps$; for example in the range of frequencies
$1.181\leq \om \leq 1.357$, the entire droplet exhibits chaotic
mixing for $\eps\geq 0.175$. For each smaller value of $\eps$ the
size reaches a maximum for a certain value $\om^m\left(\eps\right)$
of the frequency. On the one hand, this property can be used as an
optimization technique to obtain the maximal CMR size one can reach
for a given amplitude $\eps$ of the rotation. On the other hand,
$\Dt \psi$ versus $\eps$ increases quite monotonically for all
values of $\om$. The derivation of the
maxima locations and estimates of $\Dt \psi$ as functions
of the parameters and the order of resonance, will be addressed elsewhere.
The structure of the CMR in our case is rather different from
that obtained in other problems that possess resonance-induced
chaotic advection. Namely, here the size of the
CMR vanishes as $\eps$ goes to $0$ and the CMR is localized near
the resonance. In contrast, in the flow considered in, e.g.,
\cite{VWG:2007}, the mixing is caused by resonances, but the CMR
occupies a volume on the scale of the whole system. The difference
comes from the fact that the averaged change of the frequency of the
fast system vanishes in the current system, thus preventing the
trajectories starting away from the resonance from 
approaching it. This property makes the kind of flows investigated here 
useful as it may be advantageous to localize the
mixing in certain parts of the system only.
\section{\label{sec:conclu}Conclusion}
In summary, we have shown that by applying a judicious oscillatory
rotation to a translating drop (an integrable system), one can
create a chaotic mixing zone with a prescribed location and size.
The appropriate values of the parameters of the perturbation (here,
a rotation of a given frequency and amplitude) are  determined by
quantitative features of the integrable case. For any amplitude of
the rotation, the frequency optimizing the CMR size has been
obtained. Such an optimization could be useful in guiding the design
of practical mixing devices aiming at the best possible mixing rate
within individual drops.
\begin{acknowledgments}
This article is based upon work partially supported by the NSF
(grants CTS-0626070 (N.A.), CTS-0626123 (P.S.) and 0400370 (D.V.)).
D.V. is grateful to the RBRF (grant 06-01-00117) and to the Donors
of the ACS Petroleum Research Fund. C.C. acknowledges support from
Euratom-CEA (contract EUR~344-88-1~FUA~F).
\end{acknowledgments}

\end{document}